\documentclass[amsmath,amssymb,twocolumn]{revtex4-1}
\usepackage{style}
\usepackage{graphicx}  
\usepackage{dcolumn}   
\usepackage{bm}        

\begin{document}
\setcounter{totalnumber}{10}

\title{Elastic wave-turbulence and intermittency}

\author{Sergio~Chibbaro}
 \affiliation{Sorbonne Universit\'es, UPMC Univ Paris 06, UMR 7190, Institut Jean Le Rond d'Alembert, F-75005, Paris, France\\
CNRS, UMR 7190, Institut Jean Le Rond d'Alembert, F-75005, Paris, France}
\author{Christophe Josserand}
 \affiliation{Sorbonne Universit\'es, UPMC Univ Paris 06, UMR 7190, Institut Jean Le Rond d'Alembert, F-75005, Paris, France\\
CNRS, UMR 7190, Institut Jean Le Rond d'Alembert, F-75005, Paris, France}

\begin{abstract}
Weak Wave Turbulence is a powerful theory to predict statistical observables of diverse relevant physical phenomena, such as ocean waves, magnetohydrodynamics and nonlinear optics. The theory is based upon an asymptotic closure permitted in the limit of small nonlinearity. 
Here, we explore the possible deviations from this mean-field framework, in terms of anomalous scaling, focusing on the case of elastic plates. 
We establish the picture of the possible behaviors at varying the extent of nonlinearity, and we show that the mean-field theory is appropriate when all excited scales remain dominated by linear dynamics. The other picture is non-trivial and our results suggest that, when large scales contain much energy, the cascade sustains extreme events at small scales and the system displays intermittency.
\end{abstract}

\maketitle

\emph{{Introduction}}
Many phenomena can be described as random interacting waves or wave turbulence (WT)\cite{Zak,Naz,New}, notably ocean waves\cite{Hasselman,Falcon,Onorato}, capillary waves\cite{zakcap67,PusZak96,CapFauve09}  Alv\`en waves\cite{Galtier} or optical non-linear dynamics\cite{Suret}. 
At variance with hydrodynamics turbulence, the dynamics is dominated by linear waves and there is generally a clear separation of scales. Together with the large number of degrees of freedom, it allows a perturbative statistical approach for weak nonlinearities that converges asymptotically\cite{Zak,Naz,New}. For that reason, this framework is also called weak wave-turbulence (WWT) and is able to give precise predictions about statistical quantities.
Wave turbulence is basically a ``mean-field theory'' for the spectrum $n({\bf k},t)$, which neglects fluctuations and supports non-equilibrium solutions in the form of Kolmogorov-Zakharov cascade. 
In the thermodynamic limit, an asymptotic in time closure is deduced by the WWT so that the modes should be somehow close to joint gaussianity. In fact, as explained in \cite{New}, joint gaussianity is not a necessary condition for the asymptotic closure to be valid and the detailed statistics of the fluctuations
remains still a question of scientific debate.
Furthermore, anomalous scalings generally cannot be predicted from such a kind of theory. 
Since  non-linearity is indeed small in many systems of dispersive random-waves, this issue has remained scarcely considered until recently. 
An interesting analysis of a generalized 1-d nonlinear Schr\"odinger equation\cite{mmt} has shown nevertheless some discrepancies with respect to the weak wave turbulence predictions. 
This evidence has motivated the theoretical analysis of possible deviations from basic assumptions behind the theory with the possible onset of strong fluctuations, which would constitute a form of intermittency\cite{New_01a,New_01b,New_03,Lvo_04,Cho_05a,dias,Eyi_12,Galtier,meyrand2015weak}, in analogy with large deviations in turbulence\cite{Pal_07,Frisch,Fal_01}. 
Then, first evidences of such non-trivial behavior has been given for gravity waves both numerically\cite{yokoyama2004statistics} and experimentally\cite{Fal_07,Fal_10,Naz_10}. However, because of the intrinsic difficulties of gravity wave dynamics, notably steep energy transfer, crossover with capillary waves and  strong fluctuation of power injection\cite{Fal_08,Fal_10}, a detailed understanding of the origin of this
intermittent behavior is still lacking. In fact, although a potential link between the breakdown of
WWT and the appearance of intermittency has been invoked, the detailed mechanisms need still to be identified. It remains  a challenging issue of important practical interest, related to the frequency of extreme events like freak waves for instance\cite{Onorato,Janssen}. 

In this letter, we address this issue analyzing numerically the vibrations of elastic plates which can be considered now as a prototype of wave turbulence\cite{Dur_06} that is well suited for experimental investigations\cite{Mor_08,Bou_08,Miq_13,Cad_13}. 
Even though the elastic plate is a solid, for strong vibrating amplitudes, nonlinear structures have been observed, corresponding to the formation of ridges which exhibit non Gaussian flatness\cite{Miq_13}.
Experimental investigations are yet limited in the analysis of such fine structure of the dynamics, since they are typically affected by: (i) limited statistics; (ii) the effect of actual dissipation, which is hard to be modeled and seem to act over many scales\cite{Cad_13,miquel2014role}; (iii) the finite-size effect due to the vicinity of the forcing and of the measurement region\cite{miquel2011nonstationary}.
That explains why we will perform here a numerical study.

In this work, we point out a complete picture of the behaviour of the elastic wave turbulence: (i) when nonlinearity is small, WWT can be applied without compromise, and statistics of the fields are gaussian; (ii) for a sufficiently high nonlinearity, the above structures arise at large scales, in correspondence with a change in the energy spectrum;
(iii) in those cases, large fluctuations are recorded at small scales, but not at large ones, unfolding intermittency in the inertial range.

\emph{{Wave turbulence in plates}}
\begin{figure}[htbp]
\begin{center}
\includegraphics[scale=0.3]{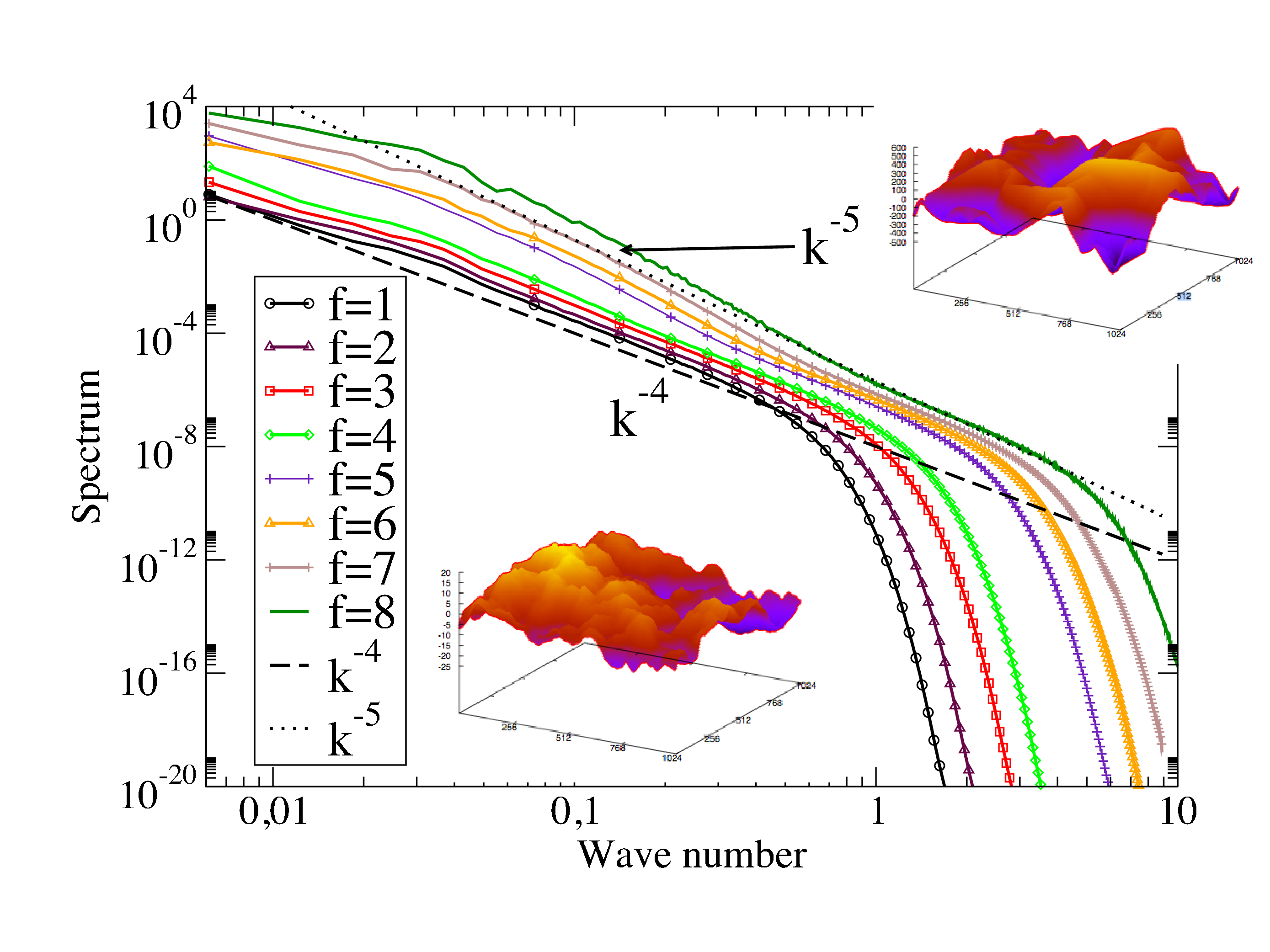} 
\caption{Wave spectrum $n_k=\lra{\vert \zeta_k \vert^2}$ for the different forcing studied.
We have used $8$ forcing amplitudes spanning two orders of magnitude (the smallest is  $10^{-6}$ and  the strongest $2\times10^{-4}$, corresponding to energy flux ranging from $4.45 \cdot 10^{-10}$ to $10^{-4}$, the energy flux being proportional to the square of the forcing amplitude). The curves correspond to increasing forcing from the bottom to the top.
The dashed lines represent the two limit curves $n_k\sim k^{-4}$ corresponding to the theoretical KZ spectrum (up to the logarithmic correction)  and $k^{-5}$ that describes the ridge formed at high forcing.
In all simulation we keep constant the ration between the plate width and the domain $\frac{h}{L}=1/2048$, in order to keep finite size effects similar.
In order to correctly describe the high forcing amplitudes, we have varied accordingly the mesh size so that our simulations have $1024\times1024$ grid points at small forcing up to $4096\times 4096$ for the highest ones. 
In the insets, surface plate deflection $\zeta(x,y)$ in the asymptotic regime for (bottom) the lowest forcing amplitude and (up) for the largest one studied in this paper. One can remark, beside the higher amplitude of the deflection, that the highest forcing creates ridge-like deformation. 
}
\label{fig1}
\end{center}
\end{figure}
Elastic vibrating plates are modeled using the dynamical version of the F\"oppl--von K\'arm\'an (FVK) equations
\cite{foppl,vk,landau} for the amplitude of deformation $\zeta(x,y,t)$ and the Airy stress function $\chi(x,y,t)$:
\begin{eqnarray} 
\rho\frac{\partial^2 \zeta}{\partial t^2} &=& - \frac{Eh^2}{12(1-\sigma^2)}\Delta^2\zeta +
\{\zeta,\chi\}  ;
\label{foppl0}\\
\frac{1}{E}\Delta^2\chi &=&- \frac{1}{2}\{\zeta,\zeta\}
\label{foppl1}
\end{eqnarray}
where  $h$ is the thickness of the elastic sheet. The material has a mass density $\rho$, a Young's 
modulus $E$ and a Poisson ratio $\sigma$.
$\Delta=\partial_{xx}+\partial_{yy}$ is the usual Laplacian and the bracket $\{\cdot,\cdot\}$ is defined by
$\{f,g\}\equiv f_{xx}g_{yy}+f_{yy}g_{xx}-2f_{xy}g_{xy},$  which is an exact divergence, so 
Eq. (\ref{foppl0}) preserves the momentum of the center of mass, namely
$\partial_{tt} \int \zeta(x,y,t) dxdy  =0$. The first term on the rhs of
(\ref{foppl0}) represents the bending while the second one
$\{\zeta,\chi\}$, a cubic nonlinearity, represents the stretching \cite{landau}. Here $\chi(x,y,t)$ is  the Airy stress function
 which follows the dynamics {\it via} eq. (\ref{foppl1}). These equations exhibit elastic waves propagating
 according to the dispersive dispersion relation $\omega_{\bf k}=\sqrt{E/12\rho(1-\sigma^2)} h k^2$,
 suggesting that the WWT formalism applies. 
This has been done in~\cite{Dur_06}: firstly, the weak nonlinearities induce a small frequency correction,
that quantifies the nonlinear interactions:
\begin{equation}
\omega^{(1)}_{\bm k}=    \frac{\pi}{ 2 h^2 }    \left[ \int_0^k \frac{\omega_q q^2}{ k^2} \lra{|\zeta_{\bm q}|^2}\, q d{q}  +  \int_k^\infty \frac{\omega_q k^2}{q^2} \lra{|\zeta_{\bm q}|^2} \, q d{q} \right] .
\label{omeg1}
\end{equation}
where $\zeta_{\bm k}$ is the Fourier transform of the displacement field $\zeta$ and the brackets $\lra{\cdot}$ indicate statistical average.
Then WWT predicts that in the absence of forcing and dissipation
the wave spectrum relaxes  towards the Rayleigh-Jeans {\it equilibrium} distribution which 
reads $\lra{|\zeta_{\bm k}|^2}= \frac{T}{\rho \omega_k^2 }$, 
where $T$ is called Temperature, by analogy with thermodynamics. 
On the other hand, additional solutions have been found corresponding to a constant flux of energy from the large scales to the small ones, the Kolmogorov-Zakharov spectrum (KZ): 
\begin{equation}
\lra{|\zeta_{\bm k}|^2}^{KZ}_{ k}= C  \frac{ P^{1/3 } \rho^{1/6}}{E^{1/2}(12(1-\sigma))^{1/6}}   \frac{\ln^{1/3}(k_*/k)}{k^4}.
\label{KZ}
\end{equation}
The logarithm correction comes from the degeneracy of the Rayleigh-Jean solutions, similarly to the
case of the nonlinear Schr\"odinger equation in 2D \cite{Dyachenko-92,malkin}.
Here $P$ is the energy flux density involved in the energy cascade (it has dimensions of mass/time$^3$), $C$ and $k_*$ being pure real numbers.\\
We solve numerically the FVK equations (\ref{foppl0},\ref{foppl1}) using a pseuso-spectral method in 
which the linear wave equation is solved exactly in the Fourier space while the nonlinear term is evaluated in the real space using Fast Fourier Transform~\cite{Dur_06}. Periodic boundary conditions 
are imposed.
Since we are interested in stationary states, we add dissipation and forcing in the equation (\ref{foppl0}).
Although, at variance with hydrodynamic turbulence, the effect of dissipation is thought to be not important in wave turbulence, realistic dissipation in plates is known to affect the spectra of vibrating plates because it is not present only at small scales~\cite{Cad_13,miquel2014role}. 
In order to avoid this problem, we have chosen to model the dissipation using a classical diffusion process $D({\bf x,t})=\gamma \Delta \zeta$, where $\gamma$ represents the relevant viscosity. It has the advantage to be mostly relevant at small scale in agreement with the WWT. Even if the large scale dissipation is less important in our model than in real plates, notice that the difference with the measured dissipation law remains reasonable\cite{miquel2011nonstationary,Cad_13}. 
The forcing is added as a white random force at large scales, as usual in wave turbulence, namely with amplitude in Fourier space following $V\Theta(k_0-k)$ where $\Theta(\cdot)$ is the Heavyside function. The main properties of the forcing are its amplitude $V$ which is varied over a very broad range of values and its 
characteristic wave length $k_0$ that we will keep constant in our simulations. The simulations start with a plate at rest and after a transient a statistically stationary regime is reached where the spectrum of the field and the energy are fluctuating around a mean value. The amplitude is 
directly related to the power injected in the system $\epsilon$ that corresponds, in this stationary 
regime, to the dissipation, following $\epsilon \propto V^2$. 
This allows us to span from very small to large excitations that can be characterized by the $\omega^{(1)}_{\bf k}/\omega_{\bf k}$ for all $k$'s. In the following, our analysis will be performed in the stationary regime for different forcing.

\emph{{Numerics and spectra} }
In fig. \ref{fig1}, we show the numerical representation of the plate deflection together with the 
corresponding displacement spectrum, for different forcing. The $k^{-4}$ slope is in line with theoretical 
predictions for all spectra at large wave numbers, whereas for strong forcing a steeper spectrum (with $| \zeta_{\bm k}|^2\sim k^{-5}$) 
appears at low wavenumbers, as recently highlighted experimentally and numerically\cite{Miq_13}.
However, the former inertial range with spectrum $\sim k^{-4}$ is still observed over at least one decade at larger $k$. Notice also that in all these spectra the logarithmic theoretical correction cannot really be
distinguished, probably because the diffusive dissipation smoothes this effect.
For high forcing, more energy is thus concentrated in coherent deformations
which have been identified to be dynamical ridge-like deformations, and
 it appears that more energy is globally absorbed by the spectrum, with the excitation of scales corresponding to high-wavenumbers. 
 
\emph{{Intermittency}}
\begin{figure}[h]
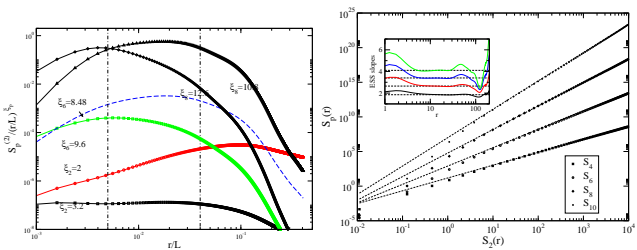

\begin{center}
\includegraphics[scale=0.15]{fig2a.eps} \includegraphics[scale=0.16]{fig2b.eps} 
\caption{(a) Structure functions of order 2,4 and 6 compensated (divided) by a scaling factor in the form $(r/L)^{\xi_p}$.
In particular they are compensated by Gaussian non-intermittent predictions (for the case shown $\xi_p=3/2p$) and by `guessed ones', for the strong forcing case 7. 
They are shown only for this case, in order to emphasize the discrepancy between Gaussian predictions and actual data. 
An inertial-range is highlighted and covers more than a decade.
The exponents used for the compensation are indicated in the graph near the corresponding curves. Given that we are dealing with the strong forcing case, we use the result $S_2\sim r^{3}$
(b) Extended self-similarity (ESS) plot of the $\zeta$ structure function $S_p$ versus $S_2$, both normalized by the value of the structure function at the
Kolmogorov scales. 
Symbols refer to the DNS data for $p=10,8, 6, 4$ from top to bottom. 
 Lines have slopes $\xi_p/\xi_2$, computed as the logarithmic local slopes of the DNS data versus distance $r$, shown in the inset. The computations give $\xi_p/\xi_2=4.05,3.4,2.65,1.85$ for $p=10,8, 6, 4$ respectively. }
\label{fig2}
\end{center}
\end{figure}
\begin{figure}[h]
\begin{center}
\includegraphics[scale=0.2]{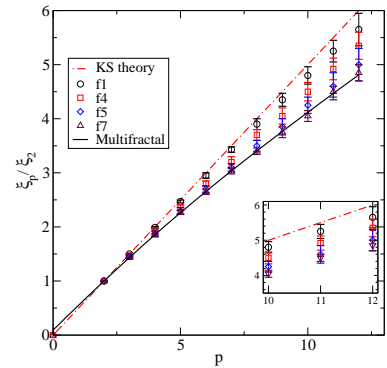} 
\caption{Structure functions exponents $\xi_p/\xi_2$ computed as in figure \ref{fig2} for different forcing.
The theoretical WWT prediction $\xi_p=\frac{p}{2}$
is shown for comparison. The rate function is concave in p.
The results obtained with multifractal random $\beta$ model are shown 
for $D_F=1$ and $x=0.65$. In the inset the detail of the higher-order exponents. 
The results obtained for the strongest forcing (f=8) collapse on those of f=7, and are not shown.
}
\label{fig3}
\end{center}
\end{figure}
To analyse intermittency and anomalous scaling, that is the lack of self-similarity, it is relevant to use the structure-functions~\cite{Monin}:
$S_p(r) = \lra{|\delta \zeta(\mathbf{x},\mathbf{r})|^p}$,
where isotropy has been used and the increment is defined as $\delta\zeta(\bf{x},\bf{r}) \equiv{\zeta(\bf{x}+\bf{r})-\zeta(\bf{x})}$. It is worth emphasizing that statistics of displacement and velocity are the same, since normal variables are a linear combination of both.
Since the spectrum $E_{\zeta}=k |\zeta|^2_{\bm k} \sim k^{-n}$, with $n\ge3$ but a possible logarithmic correction, 
the cascade is not local and Wiener-Kintchine therorem does not apply\cite{Monin}, so that the $\zeta$ field is expected to be smooth and thus $S_2(r)\sim r^2$. 
To cancel this trivial scaling, higher-order difference should be used\cite{Fal_10}, namely the second-order one, defined as $\delta\zeta^2(\bf{x},r)=\zeta(\mathbf{x}+\mathbf{r})-2\zeta(\mathbf{x})+\zeta(\mathbf{x}-\mathbf{r}) $. In this case, the KZ spectrum gives the following behavior  for the structure function of order two:
\begin{equation}
S^2_2(r) = \lra{ê\delta \zeta^2({\bf{x},\bf{r}})|^2}\sim r^{(n-1)}.
\end{equation}
At variance with hydrodynamic turbulence where there is the remarkable result
of the $4/5$ law for the correlation of third order, there is not such a result for 
the plate equations. The most fundamental result comes here from the Kolmogorov-Zakharov spectrum, so that one can deduce the correlation of second order.  In fig. \ref{fig1}, we have seen that for small forcing the spectrum is $E_{\zeta}(k)=k |\zeta|^2_k\sim k^{-3}$ and for strong forcing $E_{\zeta}(k)=k |\zeta|^2_k \sim k^{-4}$, suggesting $S_2^2\sim r^2$ and $\sim r^3$ respectively. Then, in the case of 
gaussian statistics, it would give for higher order structure functions $S_p^2(r)\sim r^p$  for low forcing (neglecting the logarithmic correction) and $S_p^2(r) \sim r^{3p/2}$ for high  forcing.
The numerical results for the structure-function of order $2$ are in line with these findings but an eventual logarithmic correction: in fact for the smallest forcing, we find $S_2^2\sim r^{\xi_2}=r^{2.2}$,
whereas for the strongest one, $S_2^2\sim r^{\xi_2}=r^{3.2}$ as it can be seen on fig. \ref{fig2}. Indeed, structure 
functions $S_p^2$ for different values of $p$ are plotted on fig. \ref{fig2}a) as function of $r$, for a high forcing case 
where the spectrum is a mixed between $k^{-5}$ at large scales and $k^{-4}$ at smaller scales. 
 For $p=2$ we observe that $S_2^2/r^{3.2}$ has an almost flat profile in the 
inertial range regime (corresponding to the region of the spectrum where $|\zeta_k|^2 \sim k^{-4}$ or $\sim k^{-5}$, that is for $r$ ranging approximately between $5$ and $100$), while $S_2^2/r^{2}$, which 
would be expected in the small forcing regime, is clearly not relevant.
Then, following this measure scaling, one would predict for a gaussian field that the high-order structure functions obey $S_p^2(r) \sim r^{3.2p/2}$ in this high forcing case.
In fig.\ref{fig2}, we compare the structure functions for $p=6$ and $p=8$ compensated by these expected scalings with those compensated using a best fit scaling law. 
We find lower exponent
values ($8.48$ instead of $9.6$ for $p=6$ and $10.8$ instead of $12.6$ for $p=8$) demonstrating that
an intermittent regime is at play. Indeed, such differences (which are above the error bars due to 
finite statistics) indicate that the dynamics cannot be understood within a single global scaling dynamics but  multiple ones are found.
Even though the simple compensation of structure functions, see fig.\ref{fig2}, seems to evidence some intermittency corrections, a thorough investigation is performed extracting the exponents 
$\xi_p$ of high-order structure functions through extended-self-similarity (ESS) technique\cite{Ben_93}, that is through the logarithmic slope of the curves obtained plotting $S_p^2(r)$ as a function of $S_2^2(r)$, as shown in fig.\ref{fig2}b. This logarithmic slope gives therefore directly the ratio $\xi_p/\xi_2$.
In figure \ref{fig3}, we show the ratio of the exponents extracted in this way for different forcing until the twelfth order (tens order only for the highest forcing because of a lack of statistics). A gaussian 
statistics would correspond to the straight line $\xi_p/\xi_2=p/2$ as plotted on the figure.
The main result is that for low-enough forcing, the results are consistent with WWT predictions, at variance with experimental findings in gravity waves\cite{Fal_10}.
\begin{figure}[h]
\begin{center}
\includegraphics[scale=0.25]{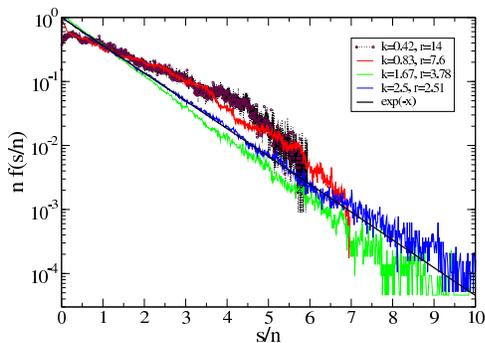} 
\caption{Pdf of the amplitude $\vert \zeta_k \vert^2$ normalized by the corresponding spectrum $n=\lra{|\zeta_{\bm k}|^2}$, for the case f=5. results for higher forcings are similar. The exponential curve represents the Rayleigh equilibrium solution, which is found for gaussian dynamics.}
\label{fig4}
\end{center}
\end{figure}
\begin{figure}[h]
\begin{center}
\includegraphics[scale=0.25]{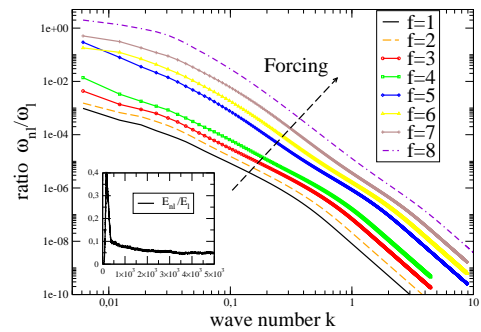} 
\caption{Nonlinear frequency with respect to linear one for al forcings used. In the inset the ratio between linear  and nonlinear energy is displayed.}
\label{fig5}
\end{center}
\end{figure}
Yet, a clear discrepancy is present for stronger forcing. 
Notably, for the strongest forcings, the discrepancy at the 12-nd order is of about 20\%, comparable with that of Eulerian hydrodynamics turbulence that is of about 30\%. Remarkably, intermittency saturates above a certain threshold in nonlinearity. 
The anomalous scaling of structure functions cannot be reproduced by a linear model, showing a non-trivial multifractal spectrum of exponents.
This suggest to model the behavior through a multiplicative process.
We have used the classical random-$\beta$ model\cite{benzi1984multifractal}. The constraint in Wave-turbulence is given by the Kolmogoroff-Zakharov spectrum. 
Using these ingredients, we obtain
$\zeta_n\sim \zeta_0 r_n^{3/2}\Pi_{j=1}^n\beta_j^{-1/2}$.
Phenomenological considerations suggest that $\beta_j=1 $ with probability $x$
and $\beta_j=B=2^{D_F-2}$,  with probability $1-x$,
where $D_F$ is the fractal dimension of the most intermittent structures.
The scaling exponents are then
$\xi_p=\frac{3}{2}-log_2\left(x+(1-x)B^{1-p/2}\right)$.
We fix $B$ considering that the minimal fractal dimension is $1$,
We have also studied the pdfs of the amplitude, which should converge to the exponential Rayleigh profile at equilibrium. 
Results displayed in fig \ref{fig4} emphasize that the equilibrium is reached in some regions at large scales, where forcing does not act, and in dissipative range, whereas in the inertial range corresponding to the $k^{-4}$ spectrum, notable deviations are found.

\emph{{Discussion} }
It is now important to question to which extent the basic hypothesis behind WWT are fulfilled.
Figure \ref{fig5} points out that the nonlinear energy remains globally much lower than the linear one, as needed by WWT asymptotic closure.
Yet, the ratio between the nonlinear and the linear frequency, which is local in spectral space, may become large when a strong forcing is imposed at large scales. 
We recognize two groups. In the first, there are those cases for which the nonlinear frequency is much smaller than the linear one for all  wavenumbers. In the second, many modes are strongly excited and the validity of WWT can be questioned in those cases, even if the modes in the inertial range remain in the WWT approximation.

In conclusion, well resolved numerical simulations of elastic turbulence show that WWT can be safely applied, provided the external excitation is sufficiently small, that is the nonlinearity is very small at all scales ($\omega^{(1)}_{\bf k}/\omega_{\bf k}<10^{-2}$).
Yet, when large scales are strongly forced, coherent structures appear, with a corresponding change in the spectrum.
While large scales remain globally gaussian, cascade sustains small scales, which in the inertial-range show intermittency in the structure functions, and the amplitude pdf deviates from Rayleigh equilibrium. 
This is consistent with the Kolmogorov cascade in which fluctuations are magnified through a multiplicative process.
From a geometrical point of view, large fluctuations are linked to the presence of tiny filaments of very small fractal dimension, corresponding here to the ridges.

\end{document}